%% file: sp_full.tex
\documentclass[11pt]{article}

\usepackage[margin=1in]{geometry}
\usepackage{amsmath,amsthm,amssymb}
\usepackage{graphicx}
\usepackage{booktabs}
\usepackage{float}
\usepackage{caption}
\usepackage{array,tabularx}
\usepackage{url}
\usepackage{hyperref}
\hypersetup{hidelinks}
\usepackage[round,authoryear]{natbib}
\usepackage{setspace}

\graphicspath{{./art/}{./results/}{./tables/}{./sim/}}
\newcommand{\E}{\mathbb{E}}
\newcommand{\Var}{\mathrm{Var}}
\newcommand{\Pp}{\mathbb{P}}
\newcommand{\R}{\mathbb{R}}
\newcommand{\bW}{\mathbf{W}}
\newcommand{\bw}{\mathbf{w}}
\newcommand{\bX}{\mathbf{X}}
\newcommand{\bbeta}{\boldsymbol{\beta}}
\newcommand{\bgamma}{\boldsymbol{\gamma}}
\newcommand{\boldeta}{\boldsymbol{\eta}}
\newcommand{\balpha}{\boldsymbol{\alpha}}

\newcommand{\bz}{\mathbf{z}}

\newtheorem{theorem}{Theorem}[section]
\newtheorem{corollary}[theorem]{Corollary}
\newtheorem{remark}[theorem]{Remark}
\newcounter{algorithm}



\newif\ifanonymous
\anonymousfalse

\title{Structured Semiparametric Estimation of Average Treatment Effects with Treatment-Specific Non-Gaussian Error Distributions}
\author{Mijeong Kim\\Department of Statistics, Ewha Womans University, Seoul 03760, South Korea\\\texttt{m.kim@ewha.ac.kr}}
\date{}

\begin{document}
\maketitle

\begin{abstract}
This paper studies average treatment effect (ATE) estimation for continuous outcomes when the treatment-covariate mean is structured but the error distributions are unknown and may differ across treatment arms in scale, skewness, or tail behavior. We introduce a semiparametric model with a finite-dimensional mean, or a prespecified basis expansion, and separate unrestricted additive error laws for treated and control outcomes. The central theoretical contribution is the ATE-efficient influence function under two treatment-specific error nuisance spaces: it combines arm-specific regression-efficient scores with variation in the marginal covariate law and is distinct from both the unrestricted AIPW gradient and a pooled location-shift score. We establish the exact nested ordering of the efficiency bounds under pooled common-error, treatment-specific-error, and unrestricted causal models, including equality conditions. A local-misspecification result further shows that the variance reduction obtained by valid pooling equals the maximal squared first-order bias incurred along unit treatment-specific directions excluded by the pooled model. We develop efficient cross-fitted plug-in and ATE-targeted implementations requiring only two one-dimensional density estimates; the targeted refinement adds a scalar update. Simulations show substantial precision gains from valid pooling under common non-Gaussian errors and protection against invalid pooling when treatment changes error shape, including under weak positivity. Applications to ACTG175 and National Supported Work data illustrate the resulting precision--robustness tradeoff for irregular continuous outcomes.
\end{abstract}

\begin{sloppypar}
\end{sloppypar}

\bigskip

\input{sp_body.tex}

\end{document}

%% file: sp_body.tex
\section{Introduction}
Modern causal-inference methodology has largely pursued robustness to nuisance-model misspecification. Doubly robust estimators, double/debiased machine-learning procedures, and newer model-averaged or multiply robust variants enlarge nuisance flexibility to protect average treatment effect (ATE) inference \citep{Chernozhukov2018DML,BachEtAl2024DoubleML,DukesEtAl2024,WangEtAl2023MR,CoulombeYang2024MR,AhrensEtAl2025}. This agenda is now central across biostatistics, medicine, and applied causal machine learning \citep{Lechner2023,BaiardiNaghi2024,Ballinari2024,CurthEtAl2024,FeuerriegelEtAl2024,AbecassisEtAl2025,TanEtAl2025}. Many continuous-outcome applications, however, face a different obstacle: the outcome distribution itself is visibly non-Gaussian. Skewed biomarkers, CD4 counts, longitudinal clinical outcomes, costs, and length-of-stay variables often motivate transformations or robust working models \citep{ChoiEtAl2022,McNeilGore1996,BuclinEtAl2011,LiuEtAl2023NonNormal,MihaylovaEtAl2011,WillifordEtAl2020}. Transformations can improve regularity, but they change the scale of the estimand and make original-scale causal interpretation depend on additional assumptions or retransformation corrections \citep{MihaylovaEtAl2011,AiNorton2000,Norton2022IHS,MullahyNorton2024}. The question addressed here is how to estimate a mean effect efficiently when a structured outcome mean is credible, Gaussian errors are not, and unrestricted nuisance learning may be unnecessarily variable.

Distributional-treatment-effect methods already allow treated and control potential-outcome distributions to differ, through quantile effects, counterfactual distribution functions, density functionals, or specified location--scale--shape families \citep{Firpo2007,ChernozhukovEtAl2013,KennedyEtAl2023,HohbergEtAl2020}. These contributions make distributional changes scientifically central, but they do not give the canonical gradient for the ATE model studied here: a structured conditional mean combined with two separate, otherwise unrestricted additive error laws.

\ifanonymous
\else
This paper studies that gap. The outcome mean is finite dimensional, or represented through a prespecified basis expansion, while the additive error law is estimated separately in each treatment arm. Treated and control errors may differ in variance, skewness, and tail behavior; only within-arm stability across baseline-covariate strata is imposed after mean adjustment. The model is therefore weaker than a common-error location shift but remains computationally parsimonious because each unknown error law is one dimensional. It is complementary to \citet{AtheyEtAl2023}: they parameterize the marginal map between potential-outcome distributions in randomized experiments, whereas we specify the covariate-indexed conditional mean and allow effect modification under either randomization or conditional exchangeability.
\fi

The contribution is model-specific and threefold. First, we derive the ATE-efficient influence function when treated and control errors generate separate nuisance tangent spaces. The resulting score uses arm-specific density scores and centering terms, then composes the regression influence function with the marginal law of $\bW$. This is not obtained by inserting two density estimates into the common-error score: the nuisance projections, information matrix, and causal composition all change. The gradient is distinct from the unrestricted AIPW gradient, the common-error regression score of \citet{Kim2023Stat}, and the marginal randomized-trial shift score of \citet{AtheyEtAl2023}.

Second, we formalize the precision--robustness frontier through an exact ordering of the efficiency bounds for the nested pooled common-error, treatment-specific-error, and unrestricted causal models, with equality characterized by tangent-space projections. A local-misspecification identity gives this ordering an operational meaning: the variance reduction obtained by pooling at a common-error distribution is exactly the maximal squared first-order bias of a pooled efficient estimator along unit treatment-specific directions that pooling excludes. Thus the smaller pooled variance and the local vulnerability to $f_0\neq f_1$ are two sides of the same projection geometry.

Third, the efficiency calculation leads to two parsimonious implementations: an efficient within-model plug-in estimator and an ATE-targeted refinement. Both require two one-dimensional density estimates rather than an unrestricted conditional outcome surface; the targeted version adds only a scalar fluctuation. Thus the paper contributes an efficiency theory, a local precision--robustness identity, and computational procedures for a regime not covered by nuisance robustness alone.

The gain is model based rather than doubly robust. Consistency requires a credible structured mean and within-arm error stability; a correct propensity-score model alone does not protect against violations of those restrictions. The intended tradeoff is explicit: when the structured outcome model is defensible, the method gains precision without imposing Gaussian errors, whereas AIPW or broader adaptive estimators remain preferable when outcome-model misspecification is the main concern.

The empirical illustrations use the ACTG175 randomized HIV trial with week-20 CD4 count and National Supported Work (NSW) data with post-intervention earnings, a wage outcome from a class of variables well known to exhibit positive skewness and long upper tails \citep{NealRosen2000}. Section~2 reviews comparator families, Sections~3--4 develop the model and estimator, Section~5 reports simulations, and Section~6 presents the applications.

\ifanonymous
\else
\section{Existing approaches to ATE estimation}
This section locates the proposed method relative to the main comparator families used in the simulations and applications. Rather than providing an exhaustive survey, the comparison below summarizes how the methods differ in the structure they exploit, the protection they seek, and the corresponding finite-sample tradeoff.

\begin{table}[H]
\centering
\caption{Conceptual comparison of the principal ATE estimator families considered in this paper.}
\label{tab:ate_method_families}
\scriptsize
\begin{tabularx}{\textwidth}{>{\raggedright\arraybackslash}p{0.17\textwidth}>{\raggedright\arraybackslash}p{0.18\textwidth}>{\raggedright\arraybackslash}p{0.26\textwidth}>{\raggedright\arraybackslash}X}
\toprule
Method family & Representative methods & Principal strategy & Main tradeoff and role in this paper \\
\midrule
Doubly robust and orthogonal & AIPW; DoubleML-IRM & Combine outcome and treatment models or use cross-fitted orthogonal scores & Protects against nuisance uncertainty, but inverse weighting and repeated nuisance learning can be variable under weak overlap \\
Targeted and multiply robust & HAL-DR; MA-DR & Enrich nuisance fits, target the ATE, or average across working specifications & Broadens robustness configurations at the cost of additional estimation complexity \\
Balancing and adaptive outcome learning & Entropy balancing; BART; forest-based TMLE & Enforce covariate balance or learn treatment-response surfaces flexibly & Useful when mean structure is uncertain, but does not exploit a credible low-dimensional outcome structure \\
Distribution structured & Firpo; Kennedy et al.; Athey et al.; Li et al. & Estimate quantiles, densities, or the marginal relation between treated and control potential-outcome distributions & Allows treatment-arm distributions to differ or exploits a credible marginal shift; closest distributional comparison with the proposed conditional-mean approach \\
\bottomrule
\end{tabularx}
\end{table}

\subsection{Nuisance-robust and adaptive estimators}
The augmented inverse-probability-weighted (AIPW) estimator is the classical doubly robust benchmark, combining outcome regression with inverse-probability correction \citep{BangRobins2005,LuncefordDavidian2004,FunkEtAl2011}. Under standard regularity and positivity conditions, it remains consistent if either the outcome regression or the propensity-score model is correctly specified. This protection against misspecification is fundamental, although limited overlap, unbalanced assignment, propensity-score miscalibration, and extreme estimated scores can produce substantial finite-sample instability \citep{KangSchafer2007,Ballinari2024,GutmanEtAl2024,XuZhangYang2026}. Its formal influence-function representation is given after the observed-data notation is introduced in Section~\ref{sec:aipw_np}.

Double/debiased machine learning extends this strategy through cross-fitting, orthogonal scores, and flexible nuisance learners \citep{Chernozhukov2018DML,BachEtAl2024DoubleML,DukesEtAl2024}. HAL-based targeted learning, model-averaged procedures, and multiply robust estimators further enlarge the range of nuisance specifications or consistency configurations considered \citep{Cefalu2016madr,WangEtAl2023MR,CoulombeYang2024MR,AhrensEtAl2025}. These methods represent the dominant response to nuisance uncertainty in modern causal machine learning \citep{BaiardiNaghi2024,Ballinari2024,Lechner2023,TanEtAl2025}.

Entropy balancing approaches the same problem through direct covariate balance, whereas BART and forest-based TMLE flexibly learn treatment-response surfaces. These are valuable when little outcome structure can be defended. Our comparison asks a different question: when a scientifically interpretable low-dimensional mean is credible, can that structure be combined with flexible non-Gaussian error modeling to improve the precision--robustness tradeoff without estimating unrestricted conditional outcome surfaces?

\subsection{Distribution-structured estimators}
The broader distributional-treatment-effect literature does not require treated and control potential outcomes to share a common distributional shape. \citet{Firpo2007} derives efficient marginal quantile-treatment-effect estimators, \citet{ChernozhukovEtAl2013} develops regression-based inference for counterfactual distributions, and \citet{KennedyEtAl2023} derives nonparametric efficiency bounds and doubly robust-style estimators for counterfactual density functionals. Distributional regression provides a complementary approach by modeling treatment effects on several parameters of a chosen location--scale--shape family \citep{HohbergEtAl2020}. These methods make $F_0\neq F_1$ central, but they target distributional features or specified distribution families rather than the ATE-efficient gradient under a structured conditional mean and two unrestricted additive error laws.

\citet{AtheyEtAl2023} develop semiparametrically efficient estimators for completely randomized experiments by placing a parametric structure directly on the marginal relation between the treatment and control potential-outcome distributions. Their leading additive specification treats the two marginal distributions as a common shift, equivalently imposing a constant marginal quantile treatment effect, and yields an efficient density-score estimator that remains well defined for thick-tailed outcomes. Their broader formulation allows other parametric transformations and, under misspecification, can retain an interpretable weighted quantile-effect target.

\citet{LiEtAl2024HeavyTailCAR} extend this approach to covariate-adaptive randomization. They establish consistency and asymptotic normality, identify possible conservativeness of the original variance estimator under more balanced designs, and develop a stratified transformed difference-in-means estimator with randomization-robust variance estimation.

The present framework places structure on a different object. It specifies a conditional outcome mean indexed by treatment and prognostic covariates while leaving two treatment-specific error laws nonparametric. It accommodates conditional-mean effect heterogeneity, does not require treated and control outcomes to share a common error shape, and targets the population ATE under either randomization or conditional exchangeability. Computationally, the marginal-shift estimator uses a marginal density score or weighted order statistics, whereas ours combines a treatment-specific regression-efficient score with a scalar ATE-targeted fluctuation.
\fi

\section{Structured model, efficient influence function, and efficiency bound}
\ifanonymous
\else
Let $O=(\bW,A,Y)$ denote one observed data unit, where $\bW\in\mathcal{W}$ is the vector of baseline covariates, $A\in\{0,1\}$ is a binary exposure, and $Y\in\R$ is the observed outcome. We observe $n$ i.i.d.\ copies $O_1,\ldots,O_n\sim P_0$, where $P_0$ denotes the true observed-data distribution, and write $\bX=(A,\bW)$. For each $a\in\{0,1\}$, let $Y(a)$ be the potential outcome under exposure level $a$, and define
\[
\mu_0(a,\bw)=\E_{P_0}(Y\mid A=a,\bW=\bw),
\qquad
g_0(a\mid\bw)=P_0(A=a\mid\bW=\bw).
\]

\subsection{Causal target and identification}
The target parameter is the average treatment effect
\[
\Psi(P_0)=\E_{P_0}\{Y(1)-Y(0)\}.
\]
We assume consistency, conditional exchangeability, and positivity:
\[
Y=Y(A),
\qquad
Y(a)\perp A\mid\bW,
\qquad
g_0(a\mid\bW)>0
\quad\text{almost surely},
\]
for each treatment level $a$ with positive target-population probability. Under these standard point-treatment assumptions \citep{RosenbaumRubin1983}, the ATE is identified by
\[
\Psi(P_0)=\E_{P_0}\{\mu_0(1,\bW)-\mu_0(0,\bW)\}.
\]
The structured model introduced next changes the efficiency calculation after identification; it does not weaken these causal assumptions or remove the need for overlap. The proposed estimator does not require a separate propensity-score model beyond identification. Propensity-score working models enter later only for comparator estimators and targeted sensitivity diagnostics.

\subsection{Treatment-specific structured outcome model}
\fi

We assume that there exists a known regression function $m:\{0,1\}\times\mathcal{W}\times\R^k\to\R$ and unknown $\bbeta_0\in\R^k$ such that
\begin{equation}\label{eq:model}
\begin{aligned}
Y&=m(A,\bW;\bbeta_0)+\varepsilon_A,\\
\varepsilon_a&\perp\bW\mid A=a,
\qquad \E(\varepsilon_a\mid A=a)=0,\\
\Var(\varepsilon_a\mid A=a)&=v_a\in(0,\infty),
\qquad a\in\{0,1\},
\end{aligned}
\end{equation}
The densities $f_{0}$ and $f_{1}$ of the two errors are otherwise unrestricted and need not be equal. Equivalently, the observed conditional outcome density is
\[
p(y\mid A=a,\bW=\bw)=f_a\{y-m(a,\bw;\bbeta_0)\}.
\]

\ifanonymous
\else
Under this model, $\mu_0(a,\bw)=m(a,\bw;\bbeta_0)$ and the ATE simplifies to
\fi
\begin{equation}\label{eq:psi_model}
\Psi(P_0)=\E_{\bW}\big[m(1,\bW;\bbeta_0)-m(0,\bW;\bbeta_0)\big].
\end{equation}
In applications, the integral in \eqref{eq:psi_model} is evaluated by the empirical distribution of $\bW$, i.e.\ by a sample average after estimating $\bbeta_0$.

The treatment-specific error laws may differ in scale, skewness, and tail behavior. The remaining restriction is that, within each treatment arm, the error distribution does not vary with baseline covariates.

\begin{corollary}[Common-error constant-shift special case]\label{cor:constant_shift}
Suppose, in addition to \eqref{eq:model}, that $f_0=f_1$ and
\[
\Delta_{\bbeta_0}(\bW)
=m(1,\bW;\bbeta_0)-m(0,\bW;\bbeta_0)
=\tau
\quad\text{almost surely}.
\]
Then, for every $\bw$ and $y$ on the relevant support,
\[
F_{1\mid\bW=\bw}(y)=F_{0\mid\bW=\bw}(y-\tau),
\]
and the marginal potential-outcome distributions satisfy
\begin{equation}\label{eq:marginal_constant_shift}
F_1(y)=F_0(y-\tau).
\end{equation}
Consequently,
\[
\Psi(P_0)=\tau.
\]
\end{corollary}
The proof is given in the Supplementary Material.

In this nested special case the mean may be parameterized as $m(A,\bW;\bbeta)=\mu(\bW;\balpha)+A\tau$. Pooling the two error samples is then valid and can improve precision. The treatment-specific procedure remains regular but generally gives up some of that pooling efficiency in return for validity when $f_0\neq f_1$.

This nested special case intersects with, but is not the target of, \citet{AtheyEtAl2023}. Their marginal location-shift formulation links the two potential-outcome distributions by a single constant shift and obtains efficiency through a common density-curvature weighting. The primary model here instead imposes structure on the conditional mean given baseline covariates: it can accommodate effect modification through $\Delta_{\bbeta_0}(\bW)$ and permits $f_0$ and $f_1$ to differ in scale, skewness, and tail behavior. Consequently, it neither assumes a single marginal shift nor admits one common curvature weight. Its efficient influence function combines arm-specific location scores with the gradient of the structured conditional mean, and the TMLE directly targets $\Psi(P_0)=\E_{\bW}\{\Delta_{\bbeta_0}(\bW)\}$.

\subsection{Regularity conditions}
We collect the standing regularity conditions here for later reference.

\noindent\textbf{(A1) Model smoothness.}
The mean function $m(A,\bW;\bbeta)$ is twice continuously differentiable in a neighborhood of $\bbeta_0$, and its gradient terms are square integrable under $P_0$.\par

\noindent\textbf{(A2) Within-arm error stability.}
For each $a$, $\varepsilon_a\perp\bW\mid A=a$, so the conditional error law within treatment arm $a$ does not vary across baseline-covariate strata. The error is centered and has strictly positive finite variance. Equality of the two arm-specific error laws is not required.\par

\noindent\textbf{(A3) Error-density regularity.}
For each $a$, the error law admits an absolutely continuous density $f_a$ that is differentiable almost everywhere. Its location score $q_a=-f_a'/f_a$ belongs to $L_2(F_a)$, so the Fisher information for location is finite; in addition, the usual boundary conditions required for the score identities and integration by parts hold. Stronger smoothness sufficient for estimating $f_a'$ at the rate used in (C3) is stated separately in the Supplementary Material.\par

\noindent\textbf{(A4) Information nonsingularity.}
The efficient information matrix $I_{\bbeta}$ for the regression parameter is nonsingular.\par

\noindent\textbf{(C1) Estimator consistency.}
For each training split, the initial estimator $\tilde\bbeta^{(-k)}$ and the arm-specific density estimators $\hat f_0^{(-k)}$ and $\hat f_1^{(-k)}$ are consistent for their population targets.\par

\noindent\textbf{(C2) Targeting accuracy.}
The targeting equation is solved to first order on each training split.\par

\noindent\textbf{(C3) Remainder control.}
The two error-density steps and their derivative estimates contribute only a second-order remainder after cross-fitting. A sufficient formulation is that, for $a=0,1$, the fold-specific nuisance errors satisfy
\[
\|\hat\bbeta^{(-k)}-\bbeta_0\|
+\|\hat f_a^{(-k)}-f_a\|_2
+\|\hat f_a^{\prime\,(-k)}-f_a'\|_2
=o_p(n^{-1/4}),
\]
uniformly over folds, together with bounded score clipping or density-floor stabilization so that the resulting second-order products are $o_p(n^{-1/2})$. The Supplementary Material records a concrete sufficient condition. In particular, for sufficiently smooth one-dimensional densities, a second-order kernel derivative estimator with derivative-oriented bandwidth can attain an $L_2$ rate faster than $n^{-1/4}$ \citep{ChaconDuong2013}.\par

\noindent\textbf{(C4) EIF moment condition.}
The resulting efficient influence function has finite second moment.\par

Conditions (A1)--(A4) are model-level assumptions; Conditions (C1)--(C4) are the additional estimator-level assumptions used in Theorem~\ref{thm:asy}.

\subsection{Interpretation and scope of the model}
Model \eqref{eq:model} is best interpreted as an arm-specific location model on a substantively chosen outcome scale. A transformation such as $\log(Y+c)$ or $\mathrm{asinh}(Y)$ can make additivity more plausible, but it should be motivated by the target estimand, interpretability, and mean-model adequacy rather than by marginal skewness alone \citep{ChoiEtAl2022}. Transformation changes the causal estimand and complicates original-scale inference \citep{AiNorton2000,Norton2022IHS,MullahyNorton2024}; conversely, retaining the natural scale is not automatically preferable when a few extreme observations make the conditional mean poorly represented or unstable. Treatment-specific error modeling reduces the need to transform merely to equalize treated and control distributions, because their scales and shapes may differ, but it does not repair a misspecified or weakly identified mean.

In observational studies, the common-error restriction $f_0=f_1$ may be particularly difficult to justify because treatment groups arise through covariate-dependent selection and may retain different unexplained scales, shapes, or tail behavior after mean adjustment. The treatment-specific formulation does not assume that $f_0\ne f_1$; rather, it avoids imposing their equality. It therefore provides a safer structured analysis when a common error law lacks strong scientific support. This flexibility does not remove confounding bias or protect against misspecification of the structured mean: conditional exchangeability, adequate overlap, and within-arm error stability remain necessary.

The mean model may be enriched through a prespecified basis of fixed dimension. The basis may contain spline terms, nonlinear transformations, and scientifically justified treatment interactions. Theorem~\ref{thm:asy}, however, treats the dimension $k$ as fixed. A growing-basis regime with $k=k_n\to\infty$, or unrestricted data-adaptive basis selection, requires additional rate, approximation-bias, and post-selection arguments and is not covered by the present theory. In practice, the basis should therefore remain modest and scientifically interpretable; if substantial flexibility is required to obtain an adequate mean, an unrestricted or machine-learning-based causal estimator is the more defensible primary analysis.

The remaining structural restriction is within-arm error stability after subtracting $m(a,\bW;\bbeta_0)$. Arm-to-arm changes in error scale or shape are allowed, but changes across covariate strata within the same arm are not. Omitted mean interactions or covariate-dependent heteroskedasticity can consequently invalidate both the efficiency calculation and its standard-error calibration. Supplementary Table~S2 makes this failure mode explicit by introducing covariate-dependent error shape while retaining the correct mean; the resulting bias and undercoverage should be read as a boundary of the method rather than as a nuisance that targeting removes.

This restriction can be screened but not certified by a finite collection of residual plots. We recommend constructing cross-fitted residuals within each arm and examining them against fitted values and prespecified important covariates, supplemented by out-of-fold prediction of squared residuals, absolute residuals, and tail indicators from $\bW$. Stratified empirical-distribution comparisons or sample-split permutation tests can provide additional screening, provided that refitting of the mean is included in their calibration. These checks remain limited in high-dimensional settings: detectable residual predictability is evidence against the model, whereas a failure to detect it is not proof of independence. When $\bW$ is too rich for credible diagnosis, or when diagnostics and flexible benchmark estimators reveal material instability, a broader heteroskedastic or nonparametric causal model should be preferred.

\subsection{EIF and efficiency bound}
This subsection derives the efficient influence function (EIF) for $\Psi(P_0)$ under model \eqref{eq:model}, building on semiparametric efficiency theory for regression with unknown error distributions \citep{BickelEtAl1998,vdVaart1998,Tsiatis2006,Kim2023Stat}.

\subsubsection{Regression-efficiency foundation for the causal derivation}
The derivation has two layers. First, the regression score must be projected separately against the nuisance tangent spaces generated by $f_0$ and $f_1$. Second, because the target is the marginal contrast in \eqref{eq:psi_model}, the causal EIF must add variation from the marginal law of $\bW$. Both layers differ from simply applying a common-error regression score to a causal contrast.

\subsubsection{EIF for the regression parameter}
For $a\in\{0,1\}$, define
\[
q_a(\bW)=\left.\frac{\partial m(a,\bW;\bbeta)}{\partial\bbeta}\right|_{\bbeta=\bbeta_0},
\qquad
\bar q_a=\E\{q_a(\bW)\mid A=a\},
\qquad
\ell_a(u)=\log f_a(u).
\]
Projecting the arm-$a$ parametric location score against the mean-zero density nuisance tangent space gives the explicit efficient score
\begin{equation}\label{eq:score_beta_proj}
S_{\mathrm{eff},\bbeta}(O)
=\sum_{a=0}^1 I(A=a)
\left[
-\ell_a'(\varepsilon_a)\{q_a(\bW)-\bar q_a\}
+\bar q_a\frac{\varepsilon_a}{v_a}
\right].
\end{equation}
The centering by $\bar q_a$ removes the component confounded with perturbations of the arm-specific error shape, while the second term restores the mean-zero normalization of that error law. No propensity-score model is required: treatment assignment affects \eqref{eq:score_beta_proj} through the empirical arm distributions and the information matrix, not through inverse-probability weights. Let
\[
I_{\bbeta}=\E\Big[S_{\mathrm{eff},\bbeta}(O)\,S_{\mathrm{eff},\bbeta}(O)^\top\Big].
\]
The absence of an explicit propensity-score denominator does not make the structured estimator immune to weak overlap. Near separation can leave parts of the treatment contrast supported mainly by mean-model extrapolation, make the arm-specific centering terms poorly estimated, and drive small eigenvalues of $I_{\bbeta}$ toward zero. Assumption (A4) excludes exact population singularity, but it does not guarantee favorable finite-sample conditioning. We therefore monitor the smallest eigenvalue and condition number of $\widehat I_{\bbeta}$ and regard severe ill-conditioning or strong disagreement with flexible benchmarks as evidence against relying on the structured analysis. The narrower claim is that \eqref{eq:score_beta_proj} avoids the additional observation-level instability created by factors such as $1/g_0(\bW_i)$; it does not remove the causal positivity requirement.
The EIF for $\bbeta$ is $D^*_{\bbeta}(O)=I_{\bbeta}^{-1}S_{\mathrm{eff},\bbeta}(O)$.
As a check, if $f_a$ is Gaussian with variance $v_a$, then $-\ell_a'(\varepsilon_a)=\varepsilon_a/v_a$ and the bracket in \eqref{eq:score_beta_proj} reduces to $q_a(\bW)\varepsilon_a/v_a$, the usual heteroskedastic Gaussian score.
Orthogonality to the tangent space of $P_{\bX}$ implies
\begin{equation}\label{eq:condmean0}
\E\{D^*_{\bbeta}(O)\mid \bX\}=0.
\end{equation}
This later removes the covariance between the marginal-$\bW$ and regression components of the causal EIF.

\subsubsection{EIF for the ATE}
Define the mean-function contrast for any $\bbeta$:
\[
\Delta_{\bbeta}(\bW)=m(1,\bW;\bbeta)-m(0,\bW;\bbeta).
\]
Then $\Psi(P)=\E\{\Delta_{\bbeta(P)}(\bW)\}$ under \eqref{eq:model}. Let
\begin{equation}\label{eq:gradpsi}
\nabla_{\bbeta}\Psi(P_0)=\E\big\{\nabla_{\bbeta} \Delta_{\bbeta_0}(\bW)\big\}.
\end{equation}

\begin{theorem}[Treatment-specific efficient influence function]\label{thm:EIF}
Under the semiparametric regression model \eqref{eq:model}, the efficient influence function for $\Psi$ at $P_0$ is
\begin{equation}\label{eq:EIF_psi}
D^*_\Psi(O)=\Big\{\Delta_{\bbeta_0}(\bW)-\Psi(P_0)\Big\} + \nabla_{\bbeta}\Psi(P_0)^\top D^*_{\bbeta}(O).
\end{equation}
\end{theorem}

Using $D^*_{\bbeta}(O)=I_{\bbeta}^{-1}S_{\mathrm{eff},\bbeta}(O)$ gives the expanded form
\begin{equation}\label{eq:EIF_psi_expanded}
\begin{aligned}
D^*_\Psi(O)
&= \Big\{\Delta_{\bbeta_0}(\bW)-\Psi(P_0)\Big\} \\
&\quad + \nabla_{\bbeta}\Psi(P_0)^\top I_{\bbeta}^{-1}
\sum_{a=0}^1 I(A=a)
\left[
-\ell_a'(\varepsilon_a)\{q_a(\bW)-\bar q_a\}
+\bar q_a\frac{\varepsilon_a}{v_a}
\right].
\end{aligned}
\end{equation}
The first term is the marginal-$\bW$ contribution; the second is the treatment-specific orthogonalized regression contribution.

Under the constant-shift submodel in Corollary~\ref{cor:constant_shift}, $\Delta_{\bbeta_0}(\bW)-\Psi(P_0)=0$ almost surely. Hence \eqref{eq:EIF_psi} reduces to
\begin{equation}\label{eq:EIF_constant_shift}
D^*_{\Psi,\mathrm{shift}}(O)
=\nabla_{\bbeta}\Psi(P_0)^\top D^*_{\bbeta}(O),
\end{equation}
and the efficiency bound contains only the regression-efficient component. Thus prognostic adjustment affects precision through estimation of the structured location parameter even though the treatment contrast itself is constant.

\subsubsection{Efficiency bound and comparison with the nonparametric model}
\begin{corollary}[Efficiency bound]\label{cor:bound}
The semiparametric efficiency bound for $\Psi$ under \eqref{eq:model} is
\[
\mathcal{I}^{-1}_{\Psi} = \Var\{\Delta_{\bbeta_0}(\bW)\} + \nabla_{\bbeta}\Psi(P_0)^\top I_{\bbeta}^{-1}\nabla_{\bbeta}\Psi(P_0).
\]
\end{corollary}

\begin{remark}
The cross-term in $\Var\{D^*_\Psi(O)\}$ vanishes because $\Delta_{\bbeta_0}(\bW)-\Psi(P_0)$ depends only on $\bW$, whereas $\E\{D^*_{\bbeta}(O)\mid \bX\}=0$ by \eqref{eq:condmean0}. Because \eqref{eq:model} is a structured submodel of the standard nonparametric causal model, its efficiency bound is no larger than the nonparametric one. The common-error submodel $f_0=f_1$ is smaller still and can have a lower bound because it permits pooling across arms.\par
\end{remark}

The preceding comparison can be stated exactly as an ordering of canonical-gradient norms. Let $\mathcal M_{\mathrm{pool}}$ denote the submodel of \eqref{eq:model} satisfying $f_0=f_1$, let $\mathcal M_{\mathrm{arm}}$ denote model~\eqref{eq:model}, and let $\mathcal M_{\mathrm{np}}$ denote the standard unrestricted observed-data causal model. For $j\in\{\mathrm{pool},\mathrm{arm},\mathrm{np}\}$, write $D_j^*$ for the canonical gradient of $\Psi$, $\overline{\mathcal T}_j$ for the closed tangent space, and $\mathcal B_j(P)=\E_P\{(D_j^*)^2\}$ for the efficiency bound.

\begin{corollary}[Nested efficiency-bound ordering]\label{cor:nested_bounds}
At every $P\in\mathcal M_{\mathrm{pool}}$ satisfying the regularity conditions,
\[
\mathcal B_{\mathrm{pool}}(P)
\leq
\mathcal B_{\mathrm{arm}}(P)
\leq
\mathcal B_{\mathrm{np}}(P).
\]
More precisely, with $\Pi_j$ denoting orthogonal projection onto $\overline{\mathcal T}_j$,
\[
D_{\mathrm{pool}}^*=\Pi_{\mathrm{pool}}D_{\mathrm{arm}}^*,
\qquad
D_{\mathrm{arm}}^*=\Pi_{\mathrm{arm}}D_{\mathrm{np}}^*,
\]
and
\begin{align*}
\mathcal B_{\mathrm{arm}}(P)-\mathcal B_{\mathrm{pool}}(P)
&=\|D_{\mathrm{arm}}^*-\Pi_{\mathrm{pool}}D_{\mathrm{arm}}^*\|_{L_2(P)}^2,\\
\mathcal B_{\mathrm{np}}(P)-\mathcal B_{\mathrm{arm}}(P)
&=\|D_{\mathrm{np}}^*-\Pi_{\mathrm{arm}}D_{\mathrm{np}}^*\|_{L_2(P)}^2.
\end{align*}
Each inequality is strict if and only if its displayed orthogonal component is nonzero.
\end{corollary}

The same projection residual also quantifies the local cost of imposing the pooled restriction when it is false. For $P_0\in\mathcal M_{\mathrm{pool}}$, let $P_{n,h}$ be a regular contiguous sequence in $\mathcal M_{\mathrm{arm}}$ through $P_0$ with score $h\in\overline{\mathcal T}_{\mathrm{arm}}$, and let $\hat\Psi_{\mathrm{pool}}$ be an efficient regular estimator in $\mathcal M_{\mathrm{pool}}$ with influence function $D_{\mathrm{pool}}^*$ at $P_0$.

\begin{corollary}[Local misspecification identity for pooling]\label{cor:local_pooling}
Under the regularity and contiguity conditions stated in the Supplementary Material,
\[
\sqrt n\{\hat\Psi_{\mathrm{pool}}-\Psi(P_{n,h})\}
\rightsquigarrow
N\!\left(
\langle D_{\mathrm{pool}}^*-D_{\mathrm{arm}}^*,h\rangle_{P_0},
\mathcal B_{\mathrm{pool}}(P_0)
\right).
\]
Moreover,
\[
\sup_{\substack{h\in\overline{\mathcal T}_{\mathrm{arm}}\cap
\overline{\mathcal T}_{\mathrm{pool}}^\perp\\
\|h\|_{L_2(P_0)}\leq1}}
\langle D_{\mathrm{pool}}^*-D_{\mathrm{arm}}^*,h\rangle_{P_0}^{,2}
=
\mathcal B_{\mathrm{arm}}(P_0)-\mathcal B_{\mathrm{pool}}(P_0).
\]
\end{corollary}

Corollary~\ref{cor:local_pooling} makes the precision--robustness tradeoff exact. At the common-error law, pooling removes the canonical-gradient component orthogonal to the pooled tangent space and gains precisely its squared norm in variance. Along local treatment-specific departures in that same direction, the discarded component reappears as first-order bias. For fixed alternatives with $f_0\neq f_1$, $\mathcal M_{\mathrm{pool}}$ is misspecified and its smaller formal variance is not an efficiency bound for the true distribution.

\subsubsection{Comparison with the usual nonparametric ATE influence function}\label{sec:aipw_np}
The structural contrast becomes clearer when we compare \eqref{eq:EIF_psi} with the usual efficient influence function for the ATE in the unrestricted causal model \citep{Hahn1998,HiranoImbensRidder2003,BangRobins2005}.
Writing $g_0(\bw)=\Pp(A=1\mid \bW=\bw)$, the standard nonparametric ATE influence function is
\begin{equation}\label{eq:aipw_eif}
\begin{aligned}
D_{\mathrm{np}}^*(O)
&= \frac{A}{g_0(\bW)}\{Y-\mu_0(1,\bW)\}
 - \frac{1-A}{1-g_0(\bW)}\{Y-\mu_0(0,\bW)\} \\
&\quad + \mu_0(1,\bW)-\mu_0(0,\bW)-\Psi(P_0).
\end{aligned}
\end{equation}
Equation~\eqref{eq:aipw_eif} is the influence-function representation underlying the AIPW estimator discussed in Section~2.1. We use AIPW as the classical doubly robust benchmark, and later compare it with broader adaptive implementations such as forest-based TMLE. Under model \eqref{eq:model}, by contrast, the response contribution enters only through the low-dimensional regression influence function $D_{\bbeta}^*(O)$, while the treatment contrast is averaged over $\bW$ through $\Delta_{\bbeta_0}(\bW)-\Psi(P_0)$. This is where a practical efficiency gain can arise: \eqref{eq:aipw_eif} carries inverse-probability weighting explicitly, so limited overlap or unstable propensity estimation can inflate variance through individual extreme weights \citep{KangSchafer2007,PetersenEtAl2012}. The structured model removes that particular amplification mechanism, but weak overlap can still degrade $I_{\bbeta}$ and increase reliance on extrapolation. The comparison is therefore between two different manifestations of limited support, not a claim that the structured estimator is uniformly robust to positivity problems.

\section{Efficient plug-in and targeted estimation}
This section gives the main cross-fitted estimator generated by the treatment-specific efficient score in \eqref{eq:score_beta_proj}. The primary construction is an efficient within-model plug-in estimator. We also report an ATE-targeted refinement in simulations and applications; it has the same first-order limit and is used as an empirical calibration of the marginal ATE score, not as a claim of uniform finite-sample RMSE improvement. The least favorable fluctuation, algorithmic details, repeated-split variance formula, and stabilized full-sample sensitivity implementation are given in the Supplementary Material.

Partition $\{1,\ldots,n\}$ into folds $\mathcal{I}_1,\ldots,\mathcal{I}_K$. On each training sample $\mathcal{I}_{-k}$, estimate the low-dimensional mean parameter by solving the full arm-specific efficient-score equation. Let $\check\bbeta^{(-k)}$ satisfy
\[
\frac{1}{|\mathcal{I}_{-k}|}
\sum_{j\in\mathcal{I}_{-k}}
\widehat S_{\mathrm{eff},\bbeta}^{(-k)}\!\big(O_j;\check\bbeta^{(-k)}\big)
= o_p(n^{-1/2}),
\]
where $\widehat S_{\mathrm{eff},\bbeta}^{(-k)}$ is formed by profiling separate one-dimensional error-density scores within the two treatment arms. The corresponding cross-fitted plug-in estimator is
\begin{equation}\label{eq:kim_plugin}
\hat\Psi_{\mathrm{plug}}
=
\frac{1}{n}\sum_{k=1}^K\sum_{i\in\mathcal{I}_k}
\Delta_{\check\bbeta^{(-k)}}(\bW_i).
\end{equation}
Because $\Psi(P)=\E\{\Delta_{\bbeta(P)}(\bW)\}$ is smooth in $\bbeta$, a delta-method expansion around $\bbeta_0$ yields
\[
\hat\Psi_{\mathrm{plug}}-\Psi(P_0)
=
\frac{1}{n}\sum_{i=1}^n
\Big[
\Delta_{\bbeta_0}(\bW_i)-\Psi(P_0)
+
\nabla_{\bbeta}\Psi(P_0)^\top D_{\bbeta}^*(O_i)
\Big]
+
o_p(n^{-1/2}),
\]
provided $\check\bbeta$ is asymptotically linear with influence function $D_{\bbeta}^*$ and the same cross-fitting remainder controls used later in Theorem~\ref{thm:asy} hold. Thus the treatment-specific plug-in benchmark and targeted estimator share the same efficient first-order limit. We retain \eqref{eq:kim_plugin} because it isolates what ATE targeting adds beyond solving the full regression score. The pooled common-error plug-in based on \citet{Kim2023Stat} is reported separately when $f_0=f_1$ is used as a working restriction.

The targeted version starts from an initial training-split fit and updates $\bbeta$ along the one-dimensional direction proportional to $\widehat I_{\bbeta}^{-1}\widehat{\nabla_{\bbeta}\Psi}$, where $\widehat I_{\bbeta}$ is the empirical information matrix for the estimated efficient regression score. The scalar update is chosen to reduce the empirical regression component of the ATE EIF. After this targeting step, aggregating the held-out predictions gives
\[
\hat\Psi_{\mathrm{cf}} = \frac{1}{n}\sum_{k=1}^K\sum_{i\in\mathcal{I}_k}\Delta_{\hat\bbeta^{(-k)}}(\bW_i).
\]
If the full efficient-score equation is already solved accurately, this update may be negligible; its practical role is to provide an observable ATE-directed calibration when regularization, numerical stopping, or stabilization leaves a residual score component.

For each observation $i$ in fold $k(i)$, compute a cross-fitted EIF estimate
\begin{equation}\label{eq:EIF_hat}
\hat D_i = \Big\{\Delta_{\hat\bbeta^{(-k(i))}}(\bW_i)-\hat\Psi_{\mathrm{cf}}\Big\} + \widehat{\nabla_{\bbeta}\Psi}^{(-k(i))\top}\, \hat D^*_{\bbeta}{}^{(-k(i))}(O_i),
\end{equation}
where $\hat D^*_{\bbeta}{}^{(-k)}(O)=\widehat{I_{\bbeta}}^{(-k)-1}\,\widehat{S_{\mathrm{eff},\bbeta}}^{(-k)}(O)$ and
\[
\widehat{\nabla_{\bbeta}\Psi}^{(-k)} = \frac{1}{|\mathcal{I}_{-k}|}\sum_{j\in\mathcal{I}_{-k}} \nabla_{\bbeta} \Delta_{\hat\bbeta^{(-k)}}(\bW_j).
\]
The standard error is estimated by
\[
\widehat{\mathrm{se}}(\hat\Psi_{\mathrm{cf}}) = \sqrt{\frac{1}{n^2}\sum_{i=1}^n \hat D_i^2},
\]
and a $95\%$ Wald confidence interval is $\hat\Psi_{\mathrm{cf}}\pm 1.96\,\widehat{\mathrm{se}}(\hat\Psi_{\mathrm{cf}})$.

Conditions (C1)--(C4) from Section~3 ensure first-order estimation of $P\mapsto \bbeta(P)$, accurate enough solving of the relevant efficient-score or targeting equation, second-order control of the error-density step, and a central limit theorem for the resulting efficient influence function.
\begin{theorem}[Asymptotic normality and efficiency]\label{thm:asy}
Under Conditions (A1)--(A4) and (C1)--(C4) from Section~3,\par
\noindent (a) $\sqrt{n}\big(\hat\Psi_{\mathrm{cf}}-\Psi(P_0)\big)\Rightarrow N\big(0,\Var\{D^*_\Psi(O)\}\big)$, and\par
\noindent (b) $\hat\Psi_{\mathrm{cf}}$ is semiparametrically efficient in model \eqref{eq:model}.\par
\end{theorem}
The Supplementary Material gives a proof sketch emphasizing the two key remainder controls: first-order asymptotic linearity of the cross-fitted regression estimator and $o_p(n^{-1/2})$ contribution of the error-density step after cross-fitting.
\section{Simulation study}\label{sec:simulation}

\subsection{Design}
The simulation study separates two questions that should not be conflated: how much precision valid error-score pooling can provide, and what is gained by allowing the error law to differ between treatment arms. Every experiment uses $n=500$ and 1,000 Monte Carlo replications.

Let $\bW=(W_1,W_2,W_3,W_4)$, where $W_1,W_2\sim N(0,1)$, $W_3\sim\mathrm{Bernoulli}(0.5)$, and $W_4\sim\mathrm{Unif}(-1,1)$ independently. The conditional mean is
\begin{equation}\label{eq:sim_mean}
\begin{aligned}
m(A,\bW;\bbeta_0)
&=A+0.8W_1-0.6W_2+0.5W_3+0.4W_4\\
&\quad+0.7AW_1-0.5AW_3.
\end{aligned}
\end{equation}
The resulting ATE is $1-0.5\E(W_3)=0.75$.

We use two treatment-assignment mechanisms. The strong-overlap mechanism is
\begin{equation}\label{eq:sim_ps_strong}
g_{\mathrm S}(\bW)
=\left[\mathrm{expit}(-0.62+0.35W_1-0.30W_2+0.30W_3-0.20W_4)\right]_{0.20}^{0.80},
\end{equation}
and the weak-positivity mechanism is
\begin{equation}\label{eq:sim_ps_weak}
g_{\mathrm W}(\bW)
=\left[\mathrm{expit}(-1.00+0.90W_1-0.80W_2+0.80W_3-0.60W_4)\right]_{0.08}^{0.92},
\end{equation}
where $[x]_l^u=\min\{u,\max(l,x)\}$. Both mechanisms have population treatment probability approximately $0.392$, so their comparison is not driven by different marginal treatment-group sizes. Their propensity-score 1st--99th percentile ranges are approximately $0.20$--$0.67$ and $0.08$--$0.92$, respectively.

Outcomes are generated as
\[
Y=m(A,\bW;\bbeta_0)+\varepsilon_A.
\]
The \emph{common-error efficiency benchmark} uses the weak-positivity mechanism and sets $\varepsilon_0\stackrel{d}{=}\varepsilon_1$. We consider either a variance-standardized $t_3$ law or the centered, variance-standardized mixture
\[
0.8N(-0.5,0.5^2)+0.2N(2,1^2).
\]
This benchmark isolates the efficiency attainable when the pooling restriction is correct. It remains difficult for generic nuisance learners because the assignment mechanism has sparse local support and the outcomes are non-Gaussian.

The \emph{treatment-specific shape benchmark} uses both assignment mechanisms. Here $\varepsilon_0$ is standardized $t_3$, whereas $\varepsilon_1$ follows the mixture above. The error variance is one in both arms, so the difference is in skewness and tail shape rather than scale. This design satisfies the primary treatment-specific model but violates the pooled restriction $f_0=f_1$. A treatment-specific scale contrast, in which the two arms share a $t_3$ shape but have different variances, is reported in the Supplementary Material.

\subsection{Estimators and implementation}
The primary comparison includes the proposed arm-specific-error TMLE and its untargeted within-model plug-in benchmark. Two pooled structured estimators impose $f_0=f_1$ and therefore serve as deliberately restricted comparators. We also include Gaussian OLS, AIPW, entropy balancing, DoubleML-IRM, HAL-DR, MA-DR, BART, and forest-based TMLE, matching the method families reviewed in Section~2.

The comparison is deliberately layered. The structured estimators and Gaussian OLS use the same correctly specified mean basis in \eqref{eq:sim_mean}. AIPW uses separate linear regressions within treatment arm on $(W_1,W_2,W_3,W_4)$, so its outcome-regression span also contains the true conditional mean. These matched-basis comparators help distinguish gains from non-Gaussian error-score estimation from gains that would arise merely by supplying the proposed method with the correct mean structure. By contrast, DoubleML-IRM, HAL-DR, BART, and forest-based TMLE receive the raw covariates and use their method-native adaptive learners; they are practical external benchmarks rather than component-wise ablations of the proposed estimator.

The arm-specific procedures estimate the two error-density scores separately; the pooled procedures estimate one common score. The proposed estimator, its plug-in benchmark, and AIPW use repeated five-fold cross-fitting. AIPW uses logistic propensity regression and the separate linear outcome regressions just described. Entropy balancing is implemented with \texttt{WeightIt}, DoubleML-IRM with \texttt{DoubleML}, HAL-DR with \texttt{hal9001}, MA-DR with \texttt{madr}, BART with \texttt{bartCause}, and forest-based TMLE with \texttt{grf}. For implementations in which we directly form inverse-probability corrections, fitted propensity scores are truncated to $[0.02,0.98]$. This numerical safeguard is wider than both generated propensity ranges, $[0.20,0.80]$ and $[0.08,0.92]$, and therefore does not alter treatment assignment or the true overlap design; it only caps occasional fitted-probability extrapolations beyond the data-generating bounds.

For each estimator we report bias, empirical standard deviation (ESD), root mean squared error (RMSE), empirical coverage of nominal 95\% Wald intervals, and average interval width. Bias and RMSE assess point estimation, whereas coverage and width jointly assess whether apparent precision is supported by valid uncertainty quantification. A narrow interval is not treated as favorable when it is achieved through material undercoverage; boldface column minima in the tables are descriptive and not an overall method ranking.

\subsection{Common-error efficiency results}
Table~\ref{tab:common_error_efficiency} reports the common-error benchmark. This is the nested setting in which pooling the one-dimensional error score is justified and should be used when supported by substantive knowledge and residual diagnostics.

\input{sim/common_error_efficiency_table.tex}

The pooled structured estimators are nearly unbiased and have the smallest RMSEs in both non-Gaussian regimes. The targeted TMLE has RMSE $0.084$ under heavy-tailed errors and $0.064$ under the skewed mixture. The corresponding RMSEs are $0.111$ and $0.112$ for Gaussian OLS, $0.144$ and $0.146$ for AIPW, and $0.171$ and $0.165$ for DoubleML-IRM. Thus AIPW and DoubleML-IRM have approximately $1.7$--$2.3$ and $2.0$--$2.6$ times the RMSE of the pooled targeted TMLE, respectively. Their average intervals are also approximately $1.7$--$2.2$ and $1.7$--$2.3$ times as wide. These results empirically corroborate, rather than establish, the efficiency-bound ordering in Corollary~\ref{cor:nested_bounds}: when the common-error restriction is correct, pooling the one-dimensional score extracts information that separate arm fits deliberately leave unused.

The most direct matched-basis comparison is the pooled plug-in versus Gaussian OLS: both use exactly \eqref{eq:sim_mean} and neither gain can be attributed to flexible mean learning, while the former replaces the Gaussian score by the estimated non-Gaussian efficient score. Under the heavy-tailed law this single change reduces RMSE from $0.111$ to $0.083$, a reduction of approximately $25\%$; under the skewed mixture it reduces RMSE from $0.112$ to $0.064$, approximately $43\%$. Because the targeted and plug-in structured results are nearly indistinguishable, these matched-basis contrasts attribute the gain, within this correctly specified design, primarily to error-score adaptation rather than to the scalar targeting step or privileged knowledge of the mean basis. The arm-specific estimator remains approximately centered but pays a visible insurance cost for allowing $f_0\ne f_1$. Relative to the pooled plug-in, its RMSE is about $30\%$ larger under the common heavy-tailed error and about $69\%$ larger under the common skewed mixture; its average interval is about $44\%$ and $79\%$ wider, respectively. The cost is therefore moderate in the first regime but material in the second, and should not be described as uniformly small. Its practical value is instead that this loss buys protection against the severe bias from invalid pooling documented in Table~\ref{tab:treatment_specific_shape}.

The pooled Wald intervals show modest undercoverage under the $t_3$ law ($0.936$ for TMLE and $0.932$ for the plug-in). These results already use repeated cross-fitting, so they should not be interpreted as a defect that an additional round of sample splitting automatically removes. The calibration diagnostic in the Supplementary Material shows that the average reported standard errors closely match the empirical standard deviations; the remaining discrepancy is mainly a finite-sample tail-normality issue under the very heavy-tailed law. We retain the conventional $1.96$ intervals rather than tune a critical value to the Monte Carlo experiment.

\subsection{Treatment-specific shape results}
Table~\ref{tab:treatment_specific_shape} reports the shape design. Here a single pooled error score is structurally incorrect because the treated and control errors have different skewness and tail behavior even after variance standardization.

\input{sim/treatment_specific_shape_overlap_table.tex}

The contrast is sharp when treatment changes error shape. The arm-specific TMLE has bias $0.020$ and $0.018$, RMSE $0.101$ and $0.112$, and coverage $0.956$ and $0.947$ under strong overlap and weak positivity, respectively. The same-model plug-in is marginally more accurate. Both estimators solve efficient first-order equations and share the same asymptotic limit; targeting is therefore not expected to provide uniform finite-sample RMSE improvement. Its narrower role is to enforce the empirical ATE-directed score equation and retain a substitution estimator explicitly calibrated to the marginal causal target. It does not repair misspecification of the structured mean or within-arm error law, and the efficient plug-in remains a legitimate simpler implementation when its score solution is stable. In contrast, invalid pooling produces biases between $-0.275$ and $-0.259$ and coverage between $0.045$ and $0.117$. Under weak positivity, AIPW and DoubleML-IRM remain approximately centered but have RMSEs $0.142$ and $0.152$, compared with $0.112$ for the arm-specific TMLE. Forest-based TMLE also becomes visibly biased, with bias $0.093$ and coverage $0.888$.

\begin{figure}[t]
\centering
\includegraphics[width=\textwidth]{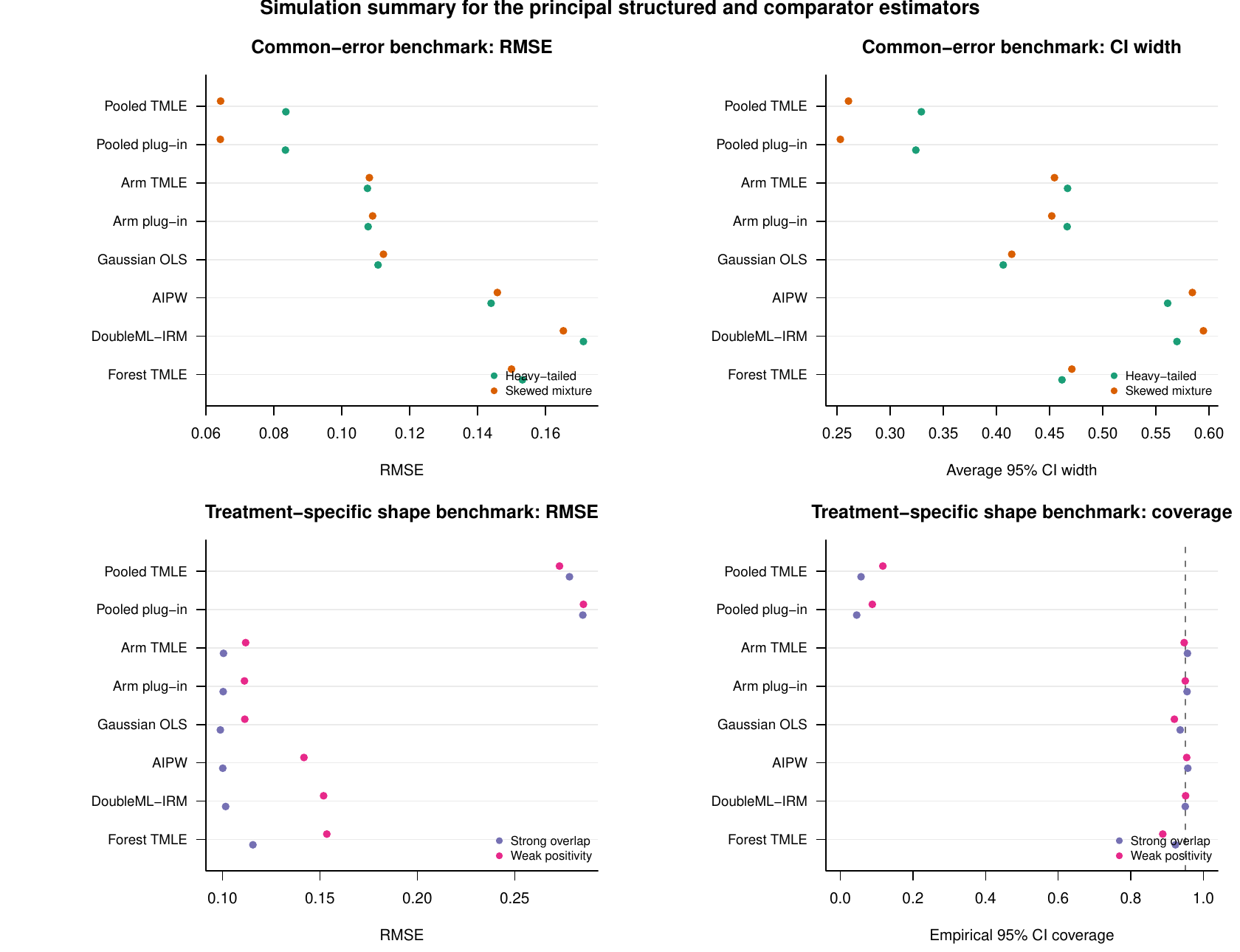}
\caption{Simulation summary for the principal structured estimators and selected comparators. The top row summarizes the common-error benchmark under weak positivity; the bottom row summarizes the treatment-specific shape benchmark under strong overlap and weak positivity. When the common-error restriction is correct, the pooled structured estimators have the smallest RMSE and shortest intervals. When treatment changes error shape, the arm-specific estimators remain stable while the invalid pooled estimators retain deceptively narrow intervals but lose coverage.}
\label{fig:simulation_summary}
\end{figure}

\subsection{Overall interpretation}
The two tables establish a precision--robustness frontier rather than universal dominance by one implementation. When the common-error restriction is credible, the pooled structured estimator converts that information into a large finite-sample precision gain. When treatment changes error shape, estimating separate arm-specific scores prevents the severe bias created by invalid pooling and remains more stable than several broad nuisance-learning alternatives under weak positivity. Overlap remains relevant even though the proposed estimator does not explicitly multiply observations by inverse propensity weights, because it determines how much information is available about each arm-specific regression component.

The practical implication is to treat pooling as an efficiency refinement, not as an automatic default. A shorter pooled interval must be evaluated together with residual diagnostics and sensitivity to arm-specific fitting. The broader comparators require less outcome-distribution structure, but do not exploit a credible low-dimensional mean together with one-dimensional non-Gaussian score information. Supplementary analyses report a finite-sample interval-calibration diagnostic, the treatment-specific scale contrast, and violations in which the error law varies with covariates within treatment arm.

\section{Empirical illustration}
Both applications use repeated five-fold cross-fitting over 10 random partitions for the primary treatment-specific analysis. We use this formulation as the primary structured analysis because it does not require equality of the two arm-specific error laws; the pooled common-error estimator is reported as an efficiency sensitivity benchmark. Supplementary Table~S4 compares the cross-fitted and stabilized full-sample implementations in ACTG175.

\begin{figure}[t]
\centering
\includegraphics[width=\textwidth]{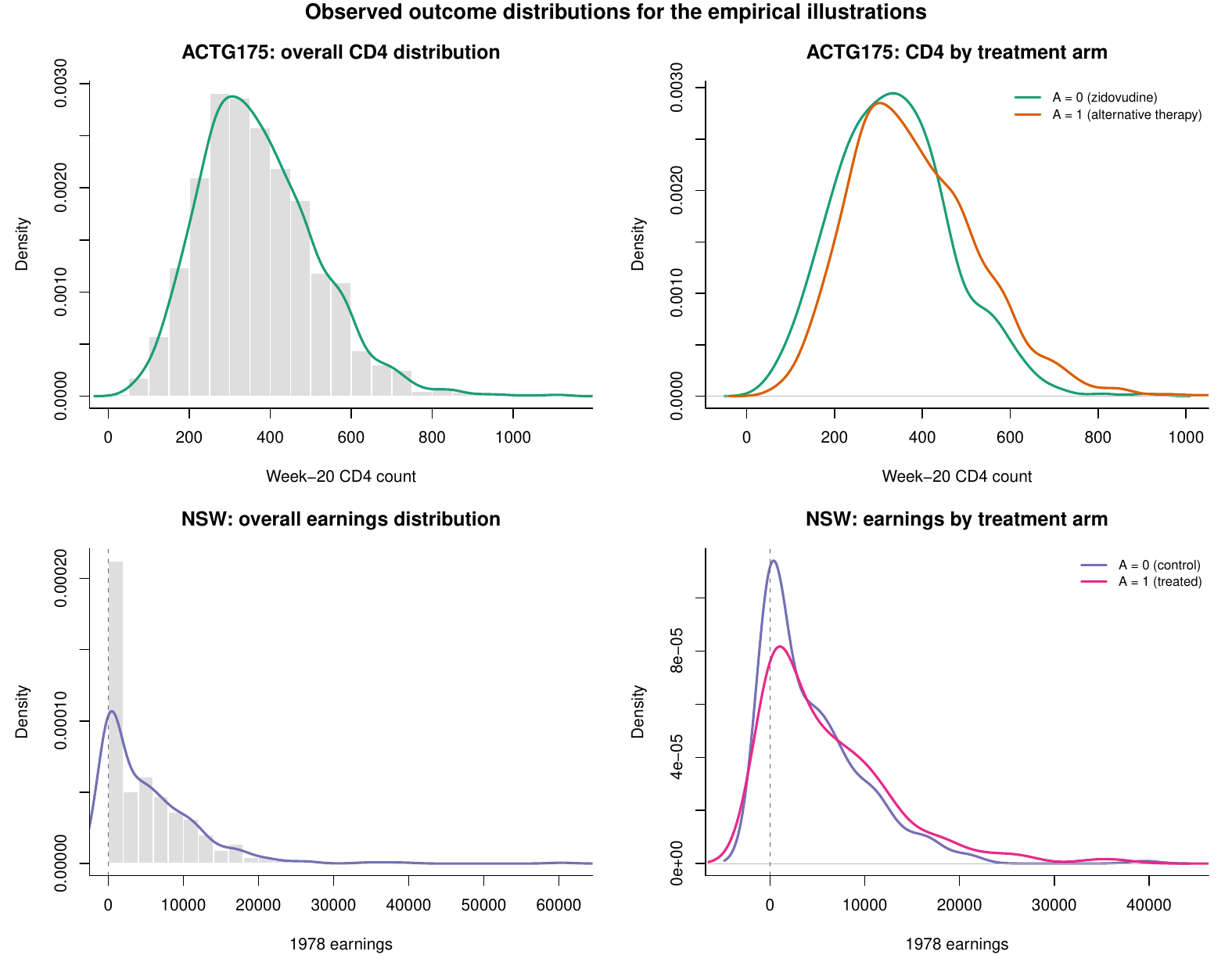}
\caption{Observed outcome distributions for the two empirical illustrations. The left column shows the overall marginal distribution of the outcome, and the right column shows treatment-stratified densities on the raw analysis scale. ACTG175 week-20 CD4 counts are visibly non-Gaussian, while NSW earnings exhibit much stronger right skewness together with substantial mass near zero. These features motivate retaining clinically or economically interpretable raw scales while allowing flexible error distributions rather than relying on Gaussian working models alone.}
\label{fig:empirical_outcome_distributions}
\end{figure}

\subsection{ACTG175 randomized benchmark}
We first consider ACTG175, a randomized HIV trial comparing zidovudine monotherapy with several alternative therapy arms \citep{HammerEtAl1996}. To align the analysis with a binary-treatment ATE, we define $A=0$ for zidovudine monotherapy and $A=1$ for the pooled alternative-therapy arms. The analysis includes all 2139 participants with complete baseline and week-20 data: 532 assigned to zidovudine monotherapy and 1607 assigned to an alternative-therapy arm. The outcome is week-20 CD4 count. This benchmark is especially relevant because the HIV literature has long recognized that raw CD4 counts are noisy and non-Gaussian enough that transformed scales such as the square-root scale are often used to obtain more regular error behavior \citep{McNeilGore1996,BuclinEtAl2011}. The same trial has also been used to estimate treatment-specific counterfactual CD4 densities and density effects \citep{KennedyEtAl2023}, underscoring that arm differences beyond the mean are scientifically meaningful. We nevertheless use raw CD4 for the main benchmark because it preserves an absolute treatment-effect interpretation in CD4 cells, and we retain log-CD4 only as a supplementary sensitivity analysis.

The structured mean model uses baseline age, weight, Karnofsky score, prior antiretroviral exposure, baseline CD4 and CD8 counts, binary risk indicators, and one treatment interaction in standardized baseline CD4. Table~\ref{tab:actg175_main} shows close point-estimate agreement. The repeated cross-fitted treatment-specific TMLE estimates $49.8$ CD4 cells with interval $(39.2,60.4)$, nearly matching its plug-in benchmark and AIPW. Its interval is approximately $8.7\%$ wider than that of the pooled common-error plug-in and $5.8\%$ wider than the AIPW interval. This modest precision cost is expected when two error scores are estimated separately in a large randomized trial; it is not evidence of finite-sample dominance by the arm-specific procedure. Supplementary Figure~S5 nevertheless shows an upper-tail arm difference in the residual Q--Q comparison. We therefore report the treatment-specific fit as the less restrictive structured analysis and the pooled fit as an efficiency sensitivity, without claiming that the observed diagnostic proves bias in the pooled estimate. Supplementary Table~S3 gives the log-CD4 sensitivity analysis, and Supplementary Table~S4 compares the cross-fitted result with the stabilized full-sample fit.

\input{results/actg175_main_table.tex}

\subsection{National Supported Work program evaluation}
To complement the biomarker benchmark with an economic outcome, we then consider National Supported Work (NSW) job-training data analyzed by \citet{Lalonde1986} and revisited by \citet{DehejiaWahba1999}. The analysis uses 445 participants with complete records: 185 program participants and 260 controls. The treatment is program participation, and the outcome is post-intervention earnings in 1978. This example is useful because the labor-earnings literature has long documented positive skewness and long upper tails in earnings distributions \citep{NealRosen2000}. In these data, 137 participants ($30.8\%$) have zero 1978 earnings, including $35.4\%$ of controls and $24.3\%$ of program participants. The outcome is therefore more accurately described as mixed discrete--continuous, with a point mass at zero and a strongly right-skewed positive component, while the baseline covariates still support a credible low-dimensional working mean specification.

We analyze 1978 earnings on their original dollar scale, $Y=\mathrm{RE78}$, with $A=\mathrm{treat}$. Baseline covariates include age, education, race indicators, marital status, no-degree status, and $\mathrm{asinh}(\mathrm{RE74})$ and $\mathrm{asinh}(\mathrm{RE75})$ for the highly skewed lagged earnings. A natural structured mean model is
\begin{equation}\label{eq:nsw_mean}
m(a,\bw;\bbeta)
=
\bbeta_0+\bbeta_1 a+\bgamma^\top \bz(\bw)+a\,\boldeta^\top \tilde{\bz}(\bw),
\end{equation}
where $\bz(\bw)$ collects the baseline main effects and $\tilde{\bz}(\bw)=\{\mathrm{asinh}(\mathrm{RE74}),\mathrm{asinh}(\mathrm{RE75})\}^\top$. The estimand is therefore the mean dollar effect, without retransformation. Because the outcome has a point mass at zero, this example does not exactly satisfy the absolute-continuity requirement in (A3); kernel density regularization supplies only a smoothed working approximation. We therefore use NSW as a deliberately difficult computational stress test, not as an empirical validation of the continuous-error efficiency theorem. A formal extension would use a two-part or hurdle formulation, modeling the probability of zero earnings and the positive continuous component separately while allowing both components to differ by treatment arm.

Table~\ref{tab:nsw_main} compares both structured models with the broad comparator set. The arm-specific-error TMLE estimates an effect of about \$1,830 with a 95\% interval from roughly \$353 to \$3,306; its width is similar to AIPW and the adaptive benchmarks. The pooled common-error plug-in gives a lower estimate near \$1,519 and a much narrower interval, but Supplementary Figure~S5 shows substantial scale and shape differences on the raw scale. The narrower pooled interval is therefore not interpreted as automatically preferable. This example illustrates the practical reason for the extension: allowing $f_0\neq f_1$ can materially change both the estimate and its uncertainty when outcome distributions differ by treatment arm.

\input{results/nsw_main_table.tex}

The split standard deviations in Table~\ref{tab:nsw_main} also reveal non-negligible partition sensitivity: they are \$102 for the arm-specific plug-in, \$167 for AIPW, and \$208 for NN-AIPW. These quantities measure variation across random partitions or repeated learner fits, not the sampling standard error of the final estimator. Repeating five-fold cross-fitting over 10 partitions reduces dependence on any single arbitrary split, while the guarded variance formula described in the Supplementary Material adds between-split variability rather than discarding it. This procedure mitigates partition noise but does not eliminate the underlying instability created by $n=445$, the point mass at zero, and the long upper tail; the reported split standard deviations are retained precisely to make that limitation visible.
\clearpage

\section{Discussion}
The principal contribution is an efficiency theory for the ATE under a structured conditional mean and two unrestricted treatment-specific additive error laws. Allowing $f_0\neq f_1$ changes the nuisance tangent space, regression-efficient score, information bound, and causal influence function. Corollary~\ref{cor:nested_bounds} identifies the efficiency gain from valid pooling as the squared norm of the arm-specific canonical-gradient component discarded by the smaller model, while Corollary~\ref{cor:local_pooling} identifies the same quantity as the maximal squared local bias along excluded treatment-specific directions. Thus the usual precision--robustness tradeoff becomes an exact local projection identity.

The simulations support this nested interpretation. When the common-error restriction is correct, pooled structured estimators convert it into substantially smaller RMSE and shorter intervals than Gaussian OLS, AIPW, DoubleML-IRM, and adaptive benchmarks. When treatment changes error shape, the arm-specific procedure avoids the bias and undercoverage produced by invalid pooling and remains competitive under weak positivity. The two findings are deliberately paired: valid structure buys precision, while relaxing the failed structure buys protection.

The method's advantage is also its boundary. It is not doubly robust: misspecification of $m(a,\bw;\bbeta)$ or covariate-dependent error structure within an arm can invalidate the ATE limit even if a propensity-score model is correct. Supplementary Table~S2 illustrates this boundary by allowing error shape to vary with $W_1$ within treatment arm; arm-specific marginal density estimation cannot replace modeling within-arm heteroskedasticity or shape variation.

For applied work, suitability should be judged by information per treatment arm rather than by a universal sample-size label. The method is most attractive when each arm can support a one-dimensional error-score estimate, the prespecified mean basis remains modest, overlap supports the modeled treatment contrasts, and the outcome is sufficiently non-Gaussian for error-score adaptation to matter. Analysts should compare arm-specific residual distributions and examine residual behavior across fitted values and important covariates. Similar arm-specific shapes support pooled estimation; visible arm differences favor treatment-specific estimation; covariate-dependent residual structure favors a broader heteroskedastic or nonparametric method.

The comparison with \citet{AtheyEtAl2023} clarifies the target. Their approach is attractive in randomized experiments when structure is naturally expressed as a marginal distributional transformation. Ours instead uses a covariate-indexed conditional mean, permits effect modification, and permits distinct arm-specific error laws. Much of the precision gain comes from this structured semiparametric model and non-Gaussian error-score adaptation, not from claiming that TMLE targeting uniformly improves finite-sample RMSE. The targeted refinement supplies ATE-directed calibration, while the plug-in remains a legitimate simpler implementation when its efficient-score solution is stable. Remaining extensions include covariate-varying error laws within arm, two-part semiparametric models for mixed discrete--continuous outcomes, post-selection inference for choosing between pooled and arm-specific fits, and targets beyond the ATE.\par

\section*{Data and Code Availability}
\ifanonymous
\else
Code for the simulations and empirical analysis is available at \url{https://github.com/mijeong-kim/structured-ate-unknown-errors}.\par
\fi

\section*{AI Use Disclosure}
During manuscript preparation, the author used OpenAI tools (ChatGPT/Codex) for language editing, limited drafting support, and code-refactoring assistance. All AI-assisted text, code, analyses, and interpretations were reviewed and verified by the author, who takes full responsibility for the final manuscript. No AI tool was used to generate primary data or to make final scientific decisions or conclusions.\par

\section*{Supplementary Material}
Proofs of Corollary~\ref{cor:constant_shift}, Theorem~\ref{thm:EIF}, Corollaries~\ref{cor:bound}, \ref{cor:nested_bounds}, and \ref{cor:local_pooling}, and Theorem~\ref{thm:asy}, together with additional simulation tables and the graphical displays omitted from the main text for space, are provided in the Supplementary Material.

\bibliography{references}

%% file: sim/common_error_efficiency_table.tex
\begin{table}[H]
\centering
\caption{Common-error efficiency benchmark at $n=500$ over 1000 Monte Carlo replications under weak positivity.}
\label{tab:common_error_efficiency}
\scriptsize
\resizebox{\textwidth}{!}{%
\begin{tabular}{lccccc@{\hspace{6pt}}ccccc}
\toprule
& \multicolumn{5}{c}{Common heavy-tailed error} & \multicolumn{5}{c}{Common skewed-mixture error} \\
\cmidrule(lr){2-6} \cmidrule(lr){7-11}
Estimator & Bias & ESD & RMSE & Cover. & Width & Bias & ESD & RMSE & Cover. & Width \\
\midrule
Pooled common-error TMLE & 0.002 & 0.084 & 0.084 & 0.936 & 0.329 & 0.001 & 0.064 & 0.064 & 0.961 & 0.261 \\
Pooled common-error plug-in (Kim) & \textbf{0.001} & \textbf{0.083} & \textbf{0.083} & 0.932 & \textbf{0.324} & \textbf{0.000} & \textbf{0.064} & \textbf{0.064} & \textbf{0.953} & \textbf{0.253} \\
Arm-specific-error TMLE & 0.004 & 0.108 & 0.108 & 0.977 & 0.467 & 0.004 & 0.108 & 0.108 & 0.965 & 0.455 \\
Arm-specific-error plug-in & 0.003 & 0.108 & 0.108 & 0.979 & 0.466 & 0.005 & 0.109 & 0.109 & 0.962 & 0.452 \\
Gaussian OLS & 0.007 & 0.110 & 0.111 & 0.935 & 0.406 & 0.002 & 0.112 & 0.112 & 0.941 & 0.414 \\
AIPW & 0.009 & 0.144 & 0.144 & 0.960 & 0.561 & 0.003 & 0.146 & 0.146 & 0.954 & 0.584 \\
Entropy balancing & 0.007 & 0.118 & 0.118 & 0.937 & 0.426 & 0.002 & 0.117 & 0.117 & 0.936 & 0.435 \\
DoubleML-IRM & 0.004 & 0.171 & 0.171 & \textbf{0.951} & 0.570 & -0.005 & 0.165 & 0.165 & 0.946 & 0.595 \\
HAL-DR & 0.020 & 0.125 & 0.127 & 0.932 & 0.444 & 0.015 & 0.128 & 0.129 & 0.932 & 0.457 \\
MA-DR & -0.026 & 0.164 & 0.166 & 0.910 & 0.554 & -0.033 & 0.170 & 0.173 & 0.890 & 0.587 \\
BART & 0.046 & 0.126 & 0.134 & 0.929 & 0.485 & 0.049 & 0.120 & 0.130 & 0.943 & 0.496 \\
Forest-based TMLE & 0.098 & 0.118 & 0.153 & 0.867 & 0.462 & 0.091 & 0.119 & 0.150 & 0.897 & 0.471 \\
\bottomrule
\end{tabular}%
}
\par\medskip
\footnotesize ESD, empirical standard deviation; Cover., empirical coverage of nominal 95\% intervals; Width, average interval width. Using the unrounded results, boldface within each error regime marks the smallest absolute bias, ESD, RMSE, and interval width, and coverage closest to 0.95. These are column-wise descriptors rather than an overall ranking, and width must be interpreted together with coverage. The common-error estimators pool density-score information across treatment arms; the arm-specific estimators estimate the two error scores separately.
\end{table}

%% file: sim/treatment_specific_shape_overlap_table.tex
\begin{table}[H]
\centering
\caption{Treatment-specific shape simulation at $n=500$ over 1000 Monte Carlo replications under strong overlap and weak positivity.}
\label{tab:treatment_specific_shape}
\scriptsize
\resizebox{\textwidth}{!}{%
\begin{tabular}{lccccc@{\hspace{6pt}}ccccc}
\toprule
& \multicolumn{5}{c}{Strong overlap} & \multicolumn{5}{c}{Weak positivity} \\
\cmidrule(lr){2-6} \cmidrule(lr){7-11}
Estimator & Bias & ESD & RMSE & Cover. & Width & Bias & ESD & RMSE & Cover. & Width \\
\midrule
Arm-specific-error TMLE & 0.020 & 0.098 & 0.101 & 0.956 & 0.413 & 0.018 & 0.110 & 0.112 & 0.947 & 0.459 \\
Arm-specific-error plug-in & 0.007 & 0.100 & 0.100 & 0.955 & 0.410 & 0.008 & 0.111 & \textbf{0.111} & \textbf{0.950} & 0.457 \\
Pooled common-error TMLE & -0.268 & \textbf{0.075} & 0.278 & 0.057 & 0.274 & -0.259 & 0.086 & 0.273 & 0.117 & 0.302 \\
Pooled common-error plug-in (Kim) & -0.275 & 0.075 & 0.285 & 0.045 & \textbf{0.268} & -0.272 & \textbf{0.086} & 0.285 & 0.088 & \textbf{0.293} \\
Gaussian OLS & -0.000 & 0.099 & \textbf{0.099} & 0.936 & 0.366 & 0.002 & 0.111 & 0.111 & 0.920 & 0.409 \\
AIPW & 0.000 & 0.100 & 0.100 & 0.957 & 0.411 & \textbf{0.000} & 0.142 & 0.142 & 0.954 & 0.576 \\
Entropy balancing & \textbf{0.000} & 0.099 & 0.099 & 0.944 & 0.388 & 0.002 & 0.117 & 0.117 & 0.928 & 0.432 \\
DoubleML-IRM & 0.000 & 0.102 & 0.102 & \textbf{0.950} & 0.401 & -0.007 & 0.152 & 0.152 & 0.951 & 0.571 \\
HAL-DR & 0.008 & 0.101 & 0.101 & 0.938 & 0.382 & 0.015 & 0.127 & 0.128 & 0.914 & 0.453 \\
MA-DR & 0.002 & 0.102 & 0.102 & 0.925 & 0.392 & -0.031 & 0.165 & 0.168 & 0.908 & 0.574 \\
BART & 0.018 & 0.105 & 0.107 & 0.954 & 0.442 & 0.045 & 0.125 & 0.132 & 0.928 & 0.492 \\
Forest-based TMLE & 0.046 & 0.106 & 0.116 & 0.923 & 0.412 & 0.093 & 0.122 & 0.154 & 0.888 & 0.468 \\
\bottomrule
\end{tabular}%
}
\par\medskip
\footnotesize ESD, empirical standard deviation; Cover., empirical coverage of nominal 95\% intervals; Width, average interval width. Using the unrounded results, boldface within each overlap regime marks the smallest absolute bias, ESD, RMSE, and interval width, and coverage closest to 0.95. These are column-wise descriptors rather than an overall ranking; in particular, the narrow pooled intervals are invalid when coverage is far below 0.95. Arm-specific estimators use separate density scores; pooled estimators impose a common error law.
\end{table}

%% file: results/actg175_main_table.tex
\begin{table}[t]
\centering
\caption{Randomized ACTG175 benchmark using week-20 CD4 count as the continuous outcome, comparing zidovudine monotherapy with the pooled alternative-therapy arms.}
\label{tab:actg175_main}
\small
\begin{tabular}{lccccc}
\toprule
Estimator & ATE & S.E. & 95\% CI & Width & Split s.d. \\
\midrule
Arm-specific-error TMLE & 49.836 & 5.404 & $(39.243,60.429)$ & 21.185 & 0.109 \\
Arm-specific-error plug-in & 49.285 & 5.417 & $(38.668,59.902)$ & 21.235 & 0.156 \\
Pooled common-error plug-in (Kim) & 49.062 & 4.971 & $(39.320,58.805)$ & 19.486 & 0.000 \\
AIPW & 49.424 & 5.106 & $(39.416,59.431)$ & 20.015 & 0.000 \\
Entropy balancing & 49.437 & 5.101 & $(39.440,59.434)$ & 19.994 & 0.000 \\
DoubleML-IRM & 49.473 & 5.181 & $(39.318,59.628)$ & 20.310 & 0.216 \\
HAL-DR & 49.711 & 5.069 & $(39.775,59.647)$ & 19.872 & 0.000 \\
MA-DR & 49.664 & 3.382 & $(43.036,56.293)$ & 13.257 & 0.000 \\
GAM-AIPW & 50.418 & 5.062 & $(40.496,60.340)$ & 19.843 & 0.000 \\
NN-AIPW & 52.531 & 5.540 & $(41.672,63.391)$ & 21.719 & 2.701 \\
BART plug-in & 50.206 & 6.688 & $(37.097,63.315)$ & 26.217 & 0.444 \\
Forest-based TMLE & 51.439 & 5.202 & $(41.244,61.635)$ & 20.391 & 0.051 \\
\bottomrule
\end{tabular}
\par\medskip
\footnotesize The trial-benchmark treatment indicator compares zidovudine monotherapy (25\% of the sample) with the pooled alternative-therapy arms (75\%). The arm-specific structured estimators use 10 repeated five-fold partitions; the common-error plug-in, AIPW, entropy balancing, HAL-DR, MA-DR, and GAM-AIPW use full-sample fits. Split s.d. tracks repeated-partition or repeated-run variability. Supplementary Table~S4 compares the cross-fitted and stabilized full-sample arm-specific fits.
\end{table}

%% file: results/nsw_main_table.tex
\begin{table}[t]
\centering
\caption{NSW empirical comparison using earnings in 1978 on the original dollar scale as the outcome.}
\label{tab:nsw_main}
\small
\begin{tabular}{lccccc}
\toprule
Estimator & ATE & S.E. & 95\% CI & Width & Split s.d. \\
\midrule
Arm-specific-error TMLE & 1829.559 & 753.418 & $(352.859,3306.258)$ & 2953.399 & 50.626 \\
Arm-specific-error plug-in & 1751.654 & 803.404 & $(176.983,3326.326)$ & 3149.343 & 102.464 \\
Common-error plug-in (Kim) & 1518.808 & 294.608 & $(941.376,2096.241)$ & 1154.865 & 35.777 \\
AIPW & 1514.639 & 722.185 & $(99.156,2930.123)$ & 2830.967 & 167.127 \\
Entropy balancing & 1470.278 & 641.545 & $(212.849,2727.706)$ & 2514.857 & 0.000 \\
DoubleML-IRM & 1662.159 & 667.890 & $(353.095,2971.223)$ & 2618.128 & 34.406 \\
HAL-DR & 1462.689 & 626.183 & $(235.370,2690.008)$ & 2454.639 & 0.000 \\
MA-DR & 1599.951 & 759.114 & $(112.087,3087.814)$ & 2975.727 & 0.000 \\
GAM-AIPW & 1463.446 & 646.503 & $(196.301,2730.592)$ & 2534.292 & 0.000 \\
NN-AIPW & 1675.754 & 720.844 & $(262.900,3088.609)$ & 2825.709 & 208.392 \\
BART plug-in & 1493.143 & 778.031 & $(-31.798,3018.085)$ & 3049.883 & 27.879 \\
Forest-based TMLE & 1616.519 & 630.697 & $(380.352,2852.685)$ & 2472.334 & 5.453 \\
\bottomrule
\end{tabular}
\par\medskip
\footnotesize The NSW illustration uses earnings in 1978 on the original dollar scale as the outcome. The arm-specific- and common-error structured estimators and AIPW are reported as averages over 10 repeated cross-fitting partitions, whereas entropy balancing, HAL-DR, MA-DR, and GAM-AIPW use full-sample nuisance fits. Split s.d. tracks repeated-splitting or repeated-run variability for the cross-fitted estimators, DoubleML-IRM, NN-AIPW, BART, and the forest-based TMLE benchmark.
\end{table}